\newcommand{\fixed}[1] { #1 }

\documentclass[fleqn]{article}
\usepackage[headings]{espcrc2}
\usepackage{graphicx}
\usepackage[figuresright]{rotating}
\newcommand{\beq}[0] { \begin{eqnarray}}
\newcommand{\eeq}[0] { \end{eqnarray}}

\newcommand{\myfig}[2] {
\begin{figure}[htb]
\vspace{9pt}
\includegraphics[width=6.5cm]{#1.eps}
\caption{#2} \label{#1}
\end{figure} }

\newcommand{\myfigv}[3]{
\begin{figure}[h]
\centerline{ \includegraphics[width=7.5cm]{#1.eps} }
\centerline{ \includegraphics[width=7.5cm]{#2.eps} }
\caption{#3} \label{#1}
\end{figure}}

\title{
The effect of hydrogen atoms on the
screw dislocation mobility in bcc iron:
A first-principles study}

\author{
M.Itakura
\address[JAEA-k]{
Center for Computational Science \& e-Systems, Japan Atomic Energy Agency.
5-1-5 Kashiwanoha, Kashiwa, Chiba 277-8587, Japan
}, 
H.Kaburaki
\address[JAEA-t]{Center for Computational Science \& e-Systems, Japan Atomic Energy Agency.
2-4 Shirakata-Shirane, Tokai-mura, Naka-gun, Ibaraki 319-1195, Japan}, 
M.Yamaguchi\addressmark[JAEA-t]
and
T.Okita
\address[race]{
Research into Artifacts, Center for Engineering, The University of Tokyo.
5-1-5 Kashiwanoha, Kashiwa, Chiba 277-8568, Japan}
}

\begin{document}

\begin{abstract}
We investigate the effect of hydrogen on the mobility of a screw dislocation in body-centered cubic (bcc) iron using first-principles calculations, and show that an increase of screw dislocation velocity is expected for a limited temperature range.
The interaction energy between a screw dislocation and hydrogen atoms is calculated for various hydrogen positions and dislocation configurations with careful estimations of the finite-size effects, and the strongest binding energy of a hydrogen atom to the stable screw dislocation configuration is estimated to be $256\pm32$ meV. 
These results are incorporated into a line tension model of a curved dislocation line to elucidate the effect of hydrogen on the dislocation migration process.
Both the softening and hardening effect of hydrogen, caused by the reduction of kink nucleation enthalpy and kink trapping, respectively, are evaluated.
A clear transition between softening and hardening behavior at the lower critical temperature is predicted, which is in qualitative agreement with experimental observation.
\vspace{1pc}

Keywords: First-principles calculation; Dislocations; Hydrogen embrittlement; Hydrogen enhanced localized plasticity
\end{abstract}

\maketitle

\section{Introduction}
Interaction between dislocations and solute atoms plays a key role in the solute hardening of metals, and in the case of body-centered cubic (bcc) metals solute softening sometimes occurs at low temperatures \cite{pink79,trinkle05}. Solute atoms either pin dislocations and hinder their motion, or reduce the Peierls barrier locally and promote kink nucleation and thus enhance dislocation motion. The precise effect of the solute atoms on plasticity depends on the core structure of the dislocation and the binding energy landscape of the solute atom around the core, and a direct investigation of such atomistic scale properties necessarily requires first-principles calculations. In the present work we use the density functional theory (DFT) calculations to investigate the effect of hydrogen on the mobility of a dislocation in bcc Fe.

Hydrogen is a special solute element because of its ubiquity in the environment and permeability into metals, resulting in a unique phenomenon known as hydrogen embrittlement (HE) in which the fracture toughness of the material is reduced significantly when subjected to a hydrogen-rich environment \cite{gangloff05}. Various mechanisms of HE have been proposed,
including hydrogen-enhanced decohesion \cite{rice89,yamaguchi12,rimoli10},
  suppression of dislocation emission by hydrogen \cite{song12} and hydrogen-enhanced localized plasticity (HELP) \cite{birnbaum94}. Each mechanism qualitatively contributes to HE, and the degree of its contribution is expected to be sensitive to the strain rate, hydrogen density and temperature. Therefore, a quantitative estimate of the contribution of each mechanism is important to identify the dominant mechanism under realistic environmental and engineering conditions and to develop a material less susceptible to HE.

Softening of materials by hydrogen (H) solute atoms has been given as a possible cause of HE in the HELP mechanism, based on the observation of the reduced flow stress \cite{matsui79} and increased screw dislocation velocity \cite{tabata83} when H atoms are introduced into bcc Fe, although hardening by H atoms also occurs depending on the subtle difference in experimental settings \cite{birnbaum94}. In bcc Fe, H atoms concentrate on the stretched region under tensile conditions \cite{wang13,ashwin08}. If concentrated H atoms induce local slips, the local strain and dislocation density increase and further concentrations of H atoms will occur, ultimately leading to the plastic instability and ductile fracture.

Since the mobility of a dislocation in bcc metals is mainly determined by that of the screw component owing to its large Peierls barrier and slow migration \cite{vitek11}, we investigate in the present work the interaction between a screw dislocation and an H atom. Although an edge dislocation strongly attracts H atoms by its long-range hydrostatic strain field and thus the effect of hydrogen on the mobility is expected to be much stronger compared to a screw dislocation, the mobility of an edge dislocation in bcc metals is very high and H atoms are likely to lower the mobility under most conditions \cite{taketomi11}. On the other hand, the mobility of a screw dislocation at low temperatures is controlled by the rate of atomic-scale kink nucleation, and a single H atom on a long screw dislocation segment can affect its nucleation rate \fixed{by lowering the nucleation energy.
This effect of hydrogen on the mobility of screw dislocations has
been proposed to explain the internal friction measurements
in iron \cite{Kirchheim4} , using a generalized framework of thermodynamics
for the defect-solute interactions \cite{Kirchheim1,Kirchheim2,Kirchheim3}. 
}

Fig. \ref{fig010cores} shows core structures of a screw dislocation for several different core positions in bcc crystals, identified by the DFT calculations \cite{woodward02,frederiksen03,clouet09,ventelon10,itakura12}.
The ``easy core'' configuration (ECC) shown in Fig. \ref{fig010cores} (a) is the most stable configuration, while the ``hard core'' configuration (HCC) shown in Fig. \ref{fig010cores} (b) is unstable or metastable and has a higher core energy owing to the large free volume inside the core. Movement of a screw dislocation in any direction requires the alternation of core structures between the ECC and HCC, resulting in the large Peierls barrier. Since H atoms are attracted to a free volume and lower the total energy, the trapping of an H atom in the HCC is expected to be stronger than that for the ECC, and the Peierls barrier is lowered by the H atom.

In the case of a screw dislocation in bcc Fe, previous DFT calculations have shown that the saddle point of a migration path between two adjacent ECCs is close to, but different from the HCC, as shown in Fig. \ref{fig010cores} (c), and that the hard core and saddle point configuration (SPC) have nearly the same core energy \cite{itakura12}.
Thus we assume that the saddle point moves toward the HCC when an H atom is in its core, and we calculate hydrogen binding energy for the ECC and HCC to investigate the effect of an H atom on the Peierls barrier.

Although the effect of an H atom to lower the Peierls barrier seems obvious when an H atom is just ahead of the screw dislocation on the slip plane, a question arises as to whether or not the average mobility is increased. Fig. \ref{fig020schem-kink} depicts the effects of an H atom on the migration process of a screw dislocation in bcc Fe. The migration is initiated by the thermal activation of a kink pair nucleus, followed by the movement of the kinks to the end of the straight screw dislocation segment, or annihilation with other kinks. For the increased average mobility, at least one H atom must always be close to and ahead of a screw dislocation line to promote the kink pair nucleation, as shown in Fig. \ref{fig020schem-kink} (a). In addition, the H atom just behind the screw dislocation line can slow or stop the kink motion and decrease the dislocation mobility, as shown in Fig. \ref{fig020schem-kink} (b). \fixed{The competition of these effects has been analyzed in detail in Ref. \cite{Kirchheim4}. In the present work, we evaluate these effects quantitatively using a line tension model which is based on DFT calculations,} and conditions for the hydrogen softening are derived.
The binding energy between hydrogen and a screw dislocation have already been calculated by DFT in Ref. \cite{zhao11}, but the estimate in Ref. \cite{zhao11} does not include the zero point energy corrections and is based on small number of $k$-point samplings, which makes it difficult to compare the results with experiments.
In the present work, we obtain more reliable estimates and compare the results with experimental observations.

The rest of this paper is organized as follows. We first describe the method used for the DFT calculations, including details of the boundary conditions. In Section 3 the results of the DFT calculations are summarized. In Section 4 the softening and hardening effect of hydrogen is estimated by the line tension model, and the results are compared with experiment. Concluding remarks are given in Section 6.

\section{Details of the DFT calculations}

The electronic structure calculations and the structure relaxations by force minimizations in the DFT steps are performed using the Vienna Ab-initio Simulation Package (VASP) \cite{vasp1,vasp2} with the projector augmented wave method and ultrasoft pseudopotentials. The exchange correlation energy is calculated by the generalized gradient approximation (GGA) with the Perdew-Burke-Ernzerhof function \cite{perdew96}. Spin-polarized calculations are employed in all cases. The Methfessel-Paxton smearing method with 0.1-eV width is used. The cutoff energy for the plane-wave basis set is 350 eV, and the convergence of hydrogen solution energy with respect to the increasing cutoff is confirmed. Structural relaxation is terminated when the maximum force acting on the movable degrees of freedom becomes less than $10$ meV/$\AA$.

The reference configurations of a screw dislocation without an H atom is obtained using a modified version of the flexible boundary method \cite{woodward02}.
It consists of DFT relaxation of atoms in the core region and Green function relaxation of atoms outside the core region. In the DFT calculations, the system is divided into two concentric hexagonal regions 1 and 2; the atoms in region 1 are relaxed while the atoms in region 2 are fixed. In the subsequent Green function relaxation, each atom's displacement relative to the linear elastic solution \cite{aniso} is calculated (denoted by $\vec{u}_i$). Then a large number of atoms are added outside region 2, and the positions of these atoms are given by the linear elastic solution. The displacement field $\vec{u}_i$ of atoms in region 2 and outside are relaxed so that the forces acting on these atoms calculated by the Hessian matrix of a perfect crystal, $\vec{f_i}=\sum_{j\neq i} M_{ij} (\vec{u_j} - \vec{u_i})$, become zero. The minimum image convention is applied for the difference of the displacement to account for the periodicity of the lattice. Displacements in region 1, which is induced by the ``core force'' \cite{clouet11}, are fixed in this step. The DFT and Green function relaxation steps are repeated until convergence. The numbers of atoms in regions 1 and 2 are 48 and 99, respectively, in the present work, as shown in Fig. \ref{fig030sys}. Throughout the present paper, Cartesian coordinates $X$, $Y$ and $Z$  parallel to $\langle \bar{2}11\rangle$, $\langle 0\bar{1}1 \rangle$ and $\langle 111 \rangle$ , respectively (see Fig. \ref{fig030sys}), are employed. The cell edge of the $Z$ direction is equal to the Burgers vector, whose length is $b$.

The two kinds of configuration ECC and HCC are used as reference configurations.
For the hexagonal supercell, $k$-points are placed on a gamma-centered mesh in the XY-plane to preserve the hexagonal symmetry and the Monkhorst-Pack $k$-point mesh is used in the Z direction. The numbers of $k$-points will be shown later for each calculation case. 
\fixed{It has been shown that
the magnetic ground state of bcc Fe, which is a uniform ferromagnetic state,
is correctly reproduced by the GGA \cite{herper99}.
Correspondingly
the bulk properties of bcc Fe are calculated using a cell containing
two Fe atoms in a ferromagnetic state, with $20^3$ $k$-points.
The lattice constant is calculated to be $a_0=2.833 \AA$, and this value is used throughout the present work.
In the dislocation calculations, 
initial magnetic moments are also set to a uniform ferromagnetic state,
while the initial moment of the H atom is set to zero, so that it
converges to the correct value regardless of the sign  of the magnetic interaction between H and Fe atoms.
It is confirmed that the magnetic moments of Fe atoms
are almost uniform in the converged state, with a modest variation near the core.
The converged magnetic moment of H atoms is negligibly small. }

After the reference state is obtained, an H atom is placed at each tetrahedral site (t-site) near the core and the atomic configurations are relaxed for each case. The hydrogen solution energy is calculated as $E_s = E_{d+H}-E_d - E_{H2}/2$, where $E_{d+H}$, $E_d$, and $E_{H2}$ denote energies of a dislocation with an H atom, a reference dislocation configuration and a hydrogen molecule, respectively. The binding energy $E_b$ of a specific site is defined as a difference in $E_s$ with respect to the bulk t-site solution energy, and is defined as positive when $E_s$ is lower than that of the bulk t-site.

The zero point energy (ZPE) correction to the solution energy is calculated from the Hessian matrix eigenvalues. We assume that Fe atoms are heavy enough compared to H atoms that the ZPE can be approximated by the motion of an H atom only. The corresponding Hessian matrix is calculated by displacing the H atom in  each of the $\pm X$, $\pm Y$ and $\pm Z$ directions by $0.015 \AA$ and observing the force acting on the H atom. ZPE correction is calculated as:
\begin{equation}
E_z=\frac{1}{2} \sum_{i=1}^3 \frac{h}{2\pi} \sqrt{k_i/m_H} ,
\end{equation}
where $k_i$ are the three eigenvalues of the $3\times 3$ Hessian matrix, $m_H$ is the mass of an H atom and $h$ is Planck's constant. We have confirmed that convergence of $E_z$ with respect to the cut-off energy, $k$-point mesh size and system size is fast and errors are of the order of a few meV, which is negligible compared to other sources of errors. The ZPE-corrected solution energy and binding energy are denoted by $E_s^Z$ and $E_b^Z$, respectively. $E_s^Z$ is given by $E_s^Z= E_s + E_z - E_{zH2}/2$, where $E_{zH2}=266$ meV is the ZPE of an H$_2$ molecule \cite{jiang04}.

In the present work, the reference dislocation configuration is replicated and three layers are stacked in the $Z$ direction to isolate the H atom from its mirror images. The number of atoms in this configuration is $341$, which is beyond the limit of feasible computation in VASP. However, not all Fe atoms are necessary to calculate the hydrogen solution energy, since the displacement of the Fe atoms caused by the hydrogen solution rapidly attenuates as the distance from the H atom increases. In the present work, Fe atoms in a hexagonal region that is centered at the H atom 
are clipped out from the reference configuration and used for the calculation of the binding energies, as shown in Fig. \ref{fig040sys2}. Outer Fe atoms are fixed to the reference configuration, because these atoms are subjected to the strong artificial force caused either by a vacuum region or a domain boundary \cite{woodward02}. In the DFT calculations of a single dislocation, the domain boundary is generally preferred over the vacuum boundary since the vacuum region induces a large amount of charge redistribution and a strong surface effect. However, if the domain boundary condition is used, $k$-point sampling in the XY-plane is required, while for the vacuum boundary one $k$-point in the XY-plane is sufficient since wave functions of any Bloch wave vector are allowed. In the present study, the domain boundary condition is employed and $3\times 3\times 8$ $k$-points are used for the high-precision calculations, while for other cases $1\times 1\times 8$ $k$-points are used. The numerical errors caused by the $k$-point sampling for the $3\times 3\times 8$ and $1\times 1\times 8$ cases are estimated as $5$ meV and $30$ meV, respectively.

Here, it is crucial to estimate the required system size for the calculation of $E_s$ with reasonably small finite-size errors.
For this purpose, the finite-size effect on $E_s$ is calculated using a cubic bcc cell with $2L^2$ Fe atoms with $L=2$, $3$ and $4$ with an H atom placed in the t-site. The Monkhorst-Pack $k$-point mesh of width $\pi/12a_0$ is used for each case. The solution energy without the relaxation of Fe atoms is $441$, $456$ and $442$ meV for $L=2$, $3$ and $4$, respectively. As has been shown in Ref. \cite{jiang04}, $E_s$ without relaxation quickly converges in the small size, so that the finite-size effect mainly comes from the relaxation. Fig. \ref{fig050fs1} (a) shows the energy change by the relaxation plotted against the inverse of the number of atoms under the constant-volume boundary conditions. Fig. \ref{fig050fs1} (a) also shows the relaxation energy calculated using the embedded atom method (EAM) potential which is fitted to various properties of hydrogen in the bcc Fe crystal calculated by DFT \cite{kimizuka11}. This EAM potential is fitted to energies without ZPE corrections and is suitable for direct comparison with DFT calculations, although for comparison with experiment, path integral molecular dynamics \cite{kimizuka11} or ZPE-corrected potential \cite{ashwin09} is required. One can see that, in both the DFT and EAM cases, the finite-size effect of the relaxation energy is inversely proportional to the system size. 


Fig. \ref{fig050fs1} (b) shows the size dependence of the total pressure induced by the H atom solution, which is also inversely proportional to the system size
(the residual pressure of the DFT case in the large volume limit is an artifact caused by a small error of the lattice constant of about $0.1\%$).
This size dependence can be explained by assuming that the displacements of the Fe atoms caused by the H atom solution are inversely proportional to the square of the distance from the H atom. $E_s$ converges to the limit $188$ meV as the system size increases while the cell volume is fixed, which is consistent with the DFT calculation with cell relaxation of $E_s=200$ meV \cite{jiang04}. From these results, a reasonable accuracy of $20$ meV is expected when the number of movable Fe atoms is about $30$. In the present work, $144$ Fe atoms in a three-layer hexagonal cell are used as a reference configuration and only the $36$ atoms in the inner hexagonal cell are relaxed, while the atoms in the outer buffer region are fixed.
The width of the buffer region is large enough so that the maximum force induced by the domain boundary on the movable inner atoms is $0.03$ eV/$\AA$  for all cases.
The finite-size effect of this hexagonal cell is directly estimated from the perfect lattice t-site solution energy using the same cell configuration shown in Fig. \ref{fig040sys2}. The outer buffer atoms are fixed and $3\times 3\times 8$ $k$-points are used, and a solution energy of $E_s=215$ meV is obtained. By comparing this with the large-volume limit $188$ meV, the finite-size error is estimated as $27$ meV. Together with the error from the $k$-point sampling, the overall precision of $E_s$ is estimated as $32$ meV and $57$ meV for $3\times 3\times 8$ and $1\times 1\times 8$ $k$-points cases, respectively. The error for $E_b$ is expected to be much smaller because of the cancellation of errors, though it is safe to employ the same error-bars for $E_b$.
\fixed{ Note that the numerical uncertainty from the exchange-correlation functional is not included in the error bars. For example, the energy of an H$_2$ molecule calculated by GGA differs from the experimental value by $85$ meV per H atom \cite{perdew96}. Therefore one should expect at least this order of error when comparing the DFT results with the experiments.
}

As mentioned in Section 1, the hydrogen binding energy of the HCC is expected to be larger than that of ECC. In DFT calculations, the hydrogen density inevitably becomes very high owing to the system size limitations and the ECC is expected to change its core structure to the HCC to gain more hydrogen binding energy at the expense of some core energy. However, to construct a model of the hydrogen-dislocation interaction, an estimate of the hydrogen binding energy in the dilute hydrogen concentration limit (where the core structure is not affected by the hydrogen solution) is required. To calculate the dilute hydrogen concentration limit, the core position is fixed to the reference configuration at a position half the thickness of the system away in the $Z$ direction from the H atom. Since the core position is related to the $Z$ displacements of three atoms surrounding the core \cite{itakura12}, the core position at a certain layer can be fixed by prohibiting relaxation in the $Z$ direction of three atoms around the core. This way, the dislocation core remains in the reference configuration and the dilute concentration limit of $E_s$ is obtained.

\section{Results}
\begin{figure*}[htb]
\centerline{ \includegraphics[width=12cm]{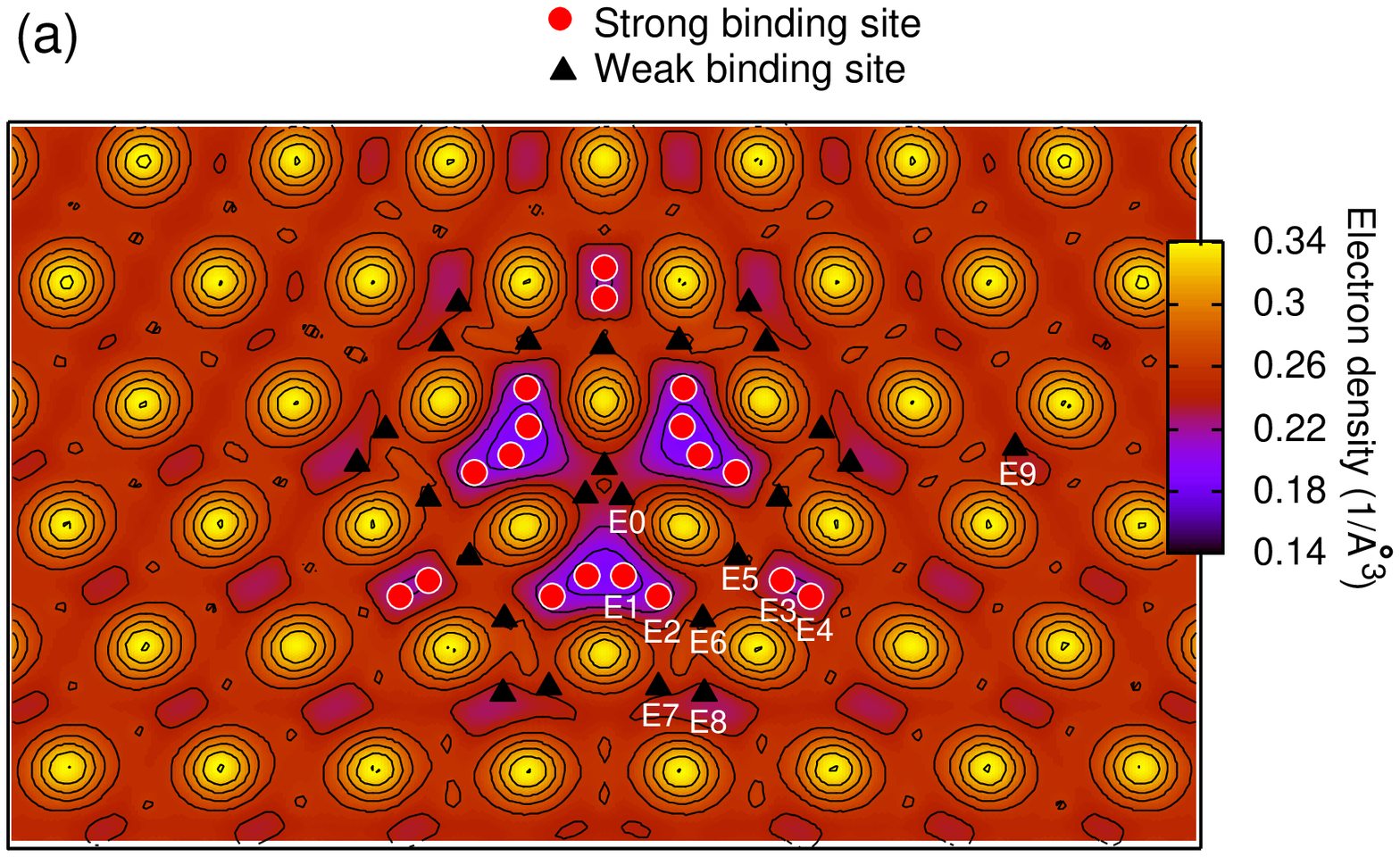} }
\centerline{ \includegraphics[width=12cm]{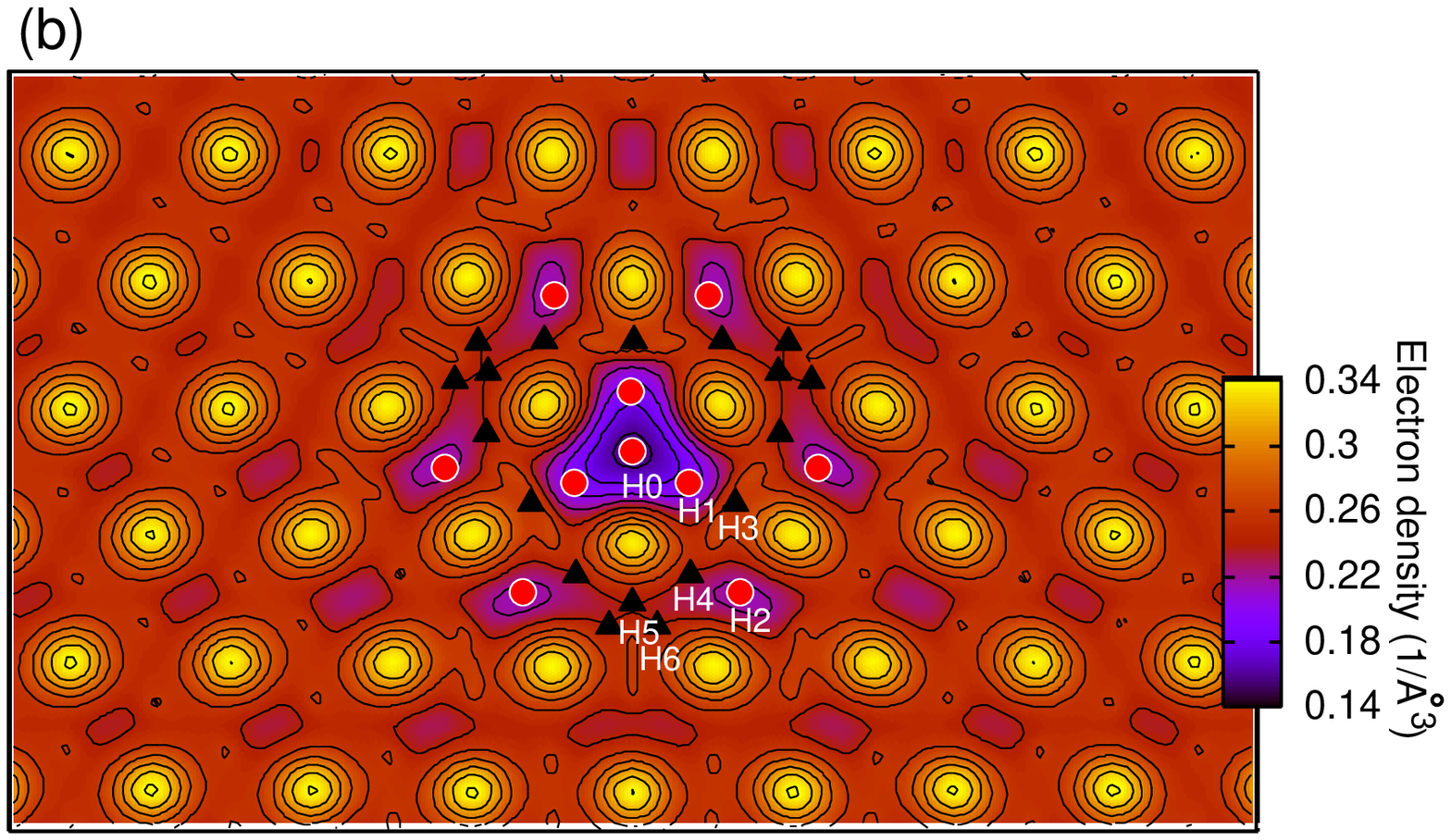} }
\caption{
Positions of the
binding sites of H atom around the core for (a) easy core and (b) hard core
configurations.
Circles and triangles represent binding sites with binding energy
larger and smaller than $100$ meV, respectively.
Labels below the symbols denote the names of the binding sites.
The minimum electron density value along the screw dislocation direction
 is also shown by the color shades and contours.
The high-density peaks correspond to the atomic rows.
} \label{fig00bind}
\end{figure*}

Fig. \ref{fig00bind} shows the positions of binding sites obtained by the DFT  calculations for ECC and HCC. The binding energies are summarized in Table \ref{tab:energy}. Fig. \ref{fig00bind} also shows minimum values of electron density along the screw dislocation line, which can be used to locate free volumes. As expected, the strong binding sites are located in the regions of free volume indicated by the low electron density. The strongest binding site for ECC is E2, which is located around the triangle adjacent to the core, and the binding energy is $E_b^Z=256\pm 32$ meV. The binding site E1 has a slightly smaller binding energy of $249$ meV, and one can see that there are three broad potential basins that encompass the binding sites E1 and E2. The four binding sites (two E1 sites and two E2 sites) in a basin are close together and the energy barriers between them are expected to be very low, and it is more appropriate to regard this basin as a single binding site which we refer to as the E1/E2 basin hereinafter. The configuration of the Fe atoms around the E1/E2 basin is shown in Fig. \ref{fig070e12}. These Fe atoms form a slanted triangular prism that has a free volume much larger than that of a perfect crystal, and the binding sites E1 and E2 are on the midplane of the prism.

It is noteworthy that this broad basin had been predicted by the EAM calculations \cite{kimizuka11}, and the barrier between the binding sites in the basin is of the order of $20$ meV in the EAM calculations. The binding energy of E2 with ZPE correction had also been calculated by two recent EAM potentials as $290$ meV for potential B in Ref. \cite{ashwin09} and $290$ meV at $300$ K in Ref. \cite{kimizuka11}, both in good agreement with the present result.

The equilibrium hydrogen concentration $C_b$ at a binding site with a binding energy $E_b^Z$ is calculated from the McLean's equation as follows:
\begin{equation}
C_b = \frac{C_0 \exp(E_b^Z/k_B T)/3}{1+C_0 \exp(E_b^Z/k_B T)/3},
\label{equ:cb}
\end{equation}
where $C_0$ is the bulk hydrogen concentration, $k_B$ is the Boltzmann constant and $T$ is the temperature. The factor $1/3$ comes from the fact that there are three t-sites per Fe atom. 

\fixed{
In Eq.  (\ref{equ:cb}), the H-H interaction is neglected.
When the interaction is attractive,
H atoms will be attracted to binding sites whose neighbour site is already occupied by an H atom,
and the binding energy becomes larger than $E_b^Z$. In this case, ignoring the H-H interaction
underestimates the concentration $C_b$.
On the other hand, when the H-H interaction is repulsive, Eq.  (\ref{equ:cb}) is valid when $C_b < 0.5$ 
since the interaction is usually only limited to the neighboring H atoms.
The precise estimate of the interaction is beyond the scope of the present work and
we use Eq. (\ref{equ:cb}) as a lower bound for $C_b$.
}

Fig. \ref{fig080cd} shows the temperature dependence of $C_b$ at the E1/E2 basin for the two cases $C_0=0.1$ atom ppm (a typical value in industrial environment) and $C_0=10$ atom ppm (a typical value in charged samples in experiments). One can see that the binding energy is strong enough to concentrate H atoms to the screw dislocation at room temperature.

The strongest binding site for HCC is H0 which is located at the center of the core, and its binding energy is $390\pm 32$ meV. This is $130$ meV stronger than the ECC case, indicating that an H atom can lower the Peierls barrier by about $130$ meV. This is consistent with the DFT calculation using the nudged elastic band method \cite{zhao11} in which binding energy difference between the ECC and SPC is reported to be about $100$ meV. The estimate of the actual reduction will be shown in the next section. It is also implied that when sufficiently large numbers of H atoms are trapped at the E1/E2 basin, the dislocation core structure may change to HCC to gain more hydrogen binding energy at the expense of the core energy. If the core structure changes to HCC when the three E1/E2 basins are occupied by H atoms with the same concentration $C_b$, H atoms in the three basins end up in the H0 site and two H2 sites. The core energy increases by $40$  meV per $b$ \cite{itakura12}, and the total hydrogen trap energy changes by an amount of $C_b(390-256)+2C_b(189-256)= 0$ meV per $b$ on average. Thus the ECC will remain unchanged regardless of the hydrogen concentration at the binding sites.

\fixed{ Fig. \ref{fig100ene-curve} shows 
the plot of the hydrogen binding energy as a function of    
the distance $r$ between the core and the H atom, for the binding sites
which are in the same $\{110\}$ planes as the core.}
The DFT data is well fitted by a Lorentzian function:
\begin{equation}
E_H(r)=\frac{390}{1+2(r/r_0)^2} \,\,\,\mbox{meV},
\end{equation}
where $r_0=\sqrt{6}a_0/3$ is the distance between two neighboring ECC positions.
Note that this function is a purely empirical one.
Fig. \ref{fig100ene-curve} also shows prediction from the linear elasticity theory \cite{wang13}.
As expected, the binding energy at the core region is much larger than that of the linear elastic interaction.

\section{Line tension model}
For the purpose of estimating the effect of H atoms on the kink pair nucleation enthalpy and kink nucleation rate, the interaction energy between the H atom and the screw dislocation obtained in the previous section is incorporated into the line tension model of a dislocation line \cite{rodney09,edagawa97}, which is expressed as an enthalpy of a curved screw dislocation configuration specified by the core positions $\vec{P}_j$ at each atomic layer $j$ of thickness $b$ as follows:
\begin{eqnarray}
E_{LT}&=&{K \over 2}\sum_j (\vec{P}_j - \vec{P}_{j+1})^2 \nonumber \\
    & &+ \sum_j E_p(\vec{P}_j) 
+\{ (\sigma*\vec{b})\times \vec{l}\}\cdot \vec{P}_j  \nonumber \\
& & -\sum_{j,k} E_H(|\vec{P}_j - \vec{P}^{H}_k|)
\label{equ:ltene} ,
\end{eqnarray}
where $K= 0.866$ eV $\AA^{-2}$ is a constant derived from the Hessian matrix of the ECC calculated by DFT \cite{itakura12}, $\vec{P}_j$ is a two-dimensional vector whose components are the $X$ and $Y$ coordinates of the dislocation core position, $E_p$ is the Peierls barrier per Burgers vector $b$, $E_H$ is the interaction energy between the dislocation line and the H atoms, and $\vec{P}^{H}_{k}$ is the  position of the $k$th H atom in the XY-plane.
The  third term is the contribution from the external stress, where $\sigma*\vec{b}$ is a tensor-vector product of the stress and Burgers vector and $\vec{l} = \vec{P}_j - \vec{P}_{j-1}$.
The inclination of the dislocation line given by $|\vec{P}_j - \vec{P}_{j+1}|/b$ is at most $1/30$ \cite{ventelon09}, and we assume that the interaction between the dislocation  and the H atoms is described well by the hydrogen binding energy of a straight screw dislocation.
Thus the interaction term $E_H$ in Eq. (\ref{equ:ltene}) is calculated only between an H atom and a representative dislocation segment which is closest to the H atom.
In addition, we assume that the position of the H atom, which is initially placed at the E1/E2 basin, remains virtually unchanged in the kink nucleation/migration process, since the binding in this site is strong throughout the process.
The two-dimensional Peierls energy $E_p$ has been estimated by the DFT calculations and fitted to a planewave expansion as follows \cite{itakura12}:
\begin{eqnarray}
E_p(P_x,P_y) =\sum_{\alpha=1}^3 \{
C_1 f_e(x_\alpha) + C_2 f_o(x_\alpha) \nonumber \\
+C_3 f_e(2x_\alpha) +C_4 f_o(2x_\alpha) \} \nonumber \\
+C_5 \left[f_e(x_1-x_2)+f_e(x_2-x_3)+f_e(x_3-x_1) \right] ,
 \label{eq:pot-fit}
 \end{eqnarray}
where
\begin{eqnarray}
f_e(x) &=& \frac{1}{2} (1-\cos 2\pi x ), \nonumber \\
f_o(x) &=& \frac{1}{2} \sin 2\pi x, \nonumber \\
x_1&=&A P_y,\nonumber \\
x_2&=&A(-P_y+\sqrt{3}P_x)/2,\nonumber \\
x_3&=&A(-P_y-\sqrt{3}P_x)/2,\nonumber \\
A&=&\sqrt{2}/a_0. \nonumber
\end{eqnarray}
The coefficients are $C_1=21.82$, $C_2=-14.51$, $C_3=2.59$, $C_4=-2.72$, and $C_5=-2.89$ meV.

The kink nucleation enthalpy with H ($E_{KH}$) and without H ($E_{K}$) for a given shear stress applied in the $[111](1\bar{1}0)$ direction is calculated using the string method \cite{wienan02}, for the dislocation migration path E2-H0-E2.
Fig. \ref{fig110ekink} shows the stress dependence of $E_{KH}$ and $E_{K}$, together with the experimental data of $E_{K}$ from Ref. \cite{spitzig70}. 
The reduction of the enthalpy, $\Delta E_K = E_{K} - E_{KH}$, is about $110$ meV for all the applied stress cases.

The discrepancy in the stress dependence of $E_K$ between the atomistic models and the experiments is a common problem for bcc metals \cite{domain05,Ta-peierls-stress,groger08}, and possible sources of this discrepancy are ascribed to several mechanisms, such as the pile-up effects \cite{groger07} and the ZPE correction \cite{proville12}.
Nevertheless, since $\Delta E_K$ only weakly depends on the stress and is mainly determined by the hydrogen binding energy, we assume that the estimate of $\Delta E_K$ is valid for all stress regions.

Fig. \ref{fig120kinksdl} (a) shows the kink nucleation/migration process in the presence of an H atom for the $400$ MPa case. The dislocation line intersects with the H atom at the highest enthalpy configuration shown by the bold line and lowers the enthalpy barrier. Fig. \ref{fig120kinksdl} (b) shows the highest enthalpy configuration seen from the $Z$ direction.  Contrary to our early expectation, the migration path of the dislocation is only slightly attracted to the HCC position from the SPC.

At the low-temperature regime, the screw dislocation velocity is determined by the kink nucleation rate given by the following Arrhenius law:
\begin{equation}
D_d N_d \exp(-E_K/k_B T),
\label{equ:kinkrate}
\end{equation}  
where $D_d$ is a prefactor and $N_d$ is the length of the dislocation in units of $b$.
From the experimental observation in Ref. \cite{caillard10}, the nucleation rate of a dislocation of length $N_d b =2 \mu$m at $300$ K under an applied shear stress of about $33$ MPa is $81$ s$^{-1}$.
The experimental value of $E_K$ for the shear stress of $33$ MPa is estimated from Fig. \ref{fig110ekink}  as $595$ meV, and the prefactor in Eq. (\ref{equ:kinkrate}) is estimated as $D_d=0.99 \times 10^8$ s$^{-1}$.

When the E1/E2 basins ahead of the dislocation in the slip direction are occupied by H atoms with concentration $C_b$, the nucleation rate is enhanced by a factor $ 1+ C_b W_k\{ \exp(\Delta E_K/k_B T)-1\}$, where $W_k \sim 10$ is the width of the kink nucleus in units of $b$.
Thus, for significant enhancement $C_b$ must be of the order of
\beq
C_b^E = \frac{1}{W_k \{\exp(\Delta E_K/k_B T)-1\}}
\label{equ:cbe}
\eeq
or greater than this value. 
As the temperature is raised, $C_b^E$ increases and $C_b$ decreases.
There is an upper critical temperature $T_U$ above which $C_b^E>C_b$ is not satisfied.
For the bulk hydrogen concentration of $0.1$ and $10$ appm, $T_U$ is $280$ and $400$ K, respectively.

For the steady enhancement of the kink nucleation, the timescale of the hydrogen diffusion should be much shorter than that of the dislocation migration.
After a dislocation migration event, the hydrogen concentration at the E1/E2 basin ahead of the dislocation (referred to as the ``promotion site'' hereinafter) is much smaller than $C_b^E$.
Enhanced nucleation does not occur until the H atoms redistribute and the hydrogen concentration at the promotion site becomes comparable with $C_b^E$.
The jump rate of the H atom at the bulk has been calculated by DFT as $D_H \exp(-E_H^m /k_B T)$, with $D_H= 5.1\times 10^{12}$ $s^{-1}$ and $E_H^m = 88$ meV \cite{ashwin08}, in reasonable agreement with the experiments \cite{jiang04}.
Since the prefactor $D_H$ is orders of magnitude greater than $D_d$, the timescale of the hydrogen redistribution is negligible if $E_{KH}$ is greater than $E_H^m$.
From the experimental values of $E_K$ in Fig. \ref{fig110ekink} and the value of $\Delta E_K=110$ meV obtained in the present work, the condition $E_{KH} > E_H^m$ sets an upper critical shear stress  $\sigma_{U1}=190$ MPa for the enhanced screw dislocation mobility, above which the hydrogen diffusion cannot catch up with the dislocation motion \cite{birnbaum94}.
We suppose that, below $\sigma_{U1}$, the hydrogen diffusion is fast enough compared to screw dislocation motion that the hydrogen density at the E1/E2 basin is approximated well by $C_b$ in Eq. (\ref{equ:cb}).
For more precise evaluation, a kinetic Monte Carlo simulation of combined dislocation motion and hydrogen diffusion is required.

Next, we investigate the kink trapping effect of the H atom depicted in Fig. \ref{fig150slow}.
Suppose that an H atom is at an E1/E2 basin behind the dislocation line, and a kink is moving from left to right. \fixed{Note that the kink migration is orders of magnitude faster than the hydrogen diffusion and the H atom remains at the basin throughout the kink migration process.} After the kink sweeps past the H atom, the relative position of H atom to the dislocation core changes from the E1/E2 basin to E8, and its solution energy increases by $179\pm 57$ meV.
This kink trapping barrier, denoted by $E_t$, is reduced as the applied shear stress $\sigma$ increases because the enthalpy as a function of the kink position $z$ has a term $-z\sigma b h$, where $h =\sqrt{6}a_0/3$ is the kink height, and there is a critical stress $\sigma_c$ at which the barrier vanishes, as shown in Fig. \ref{fig150slow} (b). 
The values of $E_t$ for various applied shear stress are calculated using the string method, and are plotted against the shear stress in Fig. \ref{fig160et}, together with the plots of $E_{K}$ obtained by the DFT calculations and experiments.
The estimated value of $\sigma_c$ is $265$ MPa.

Below $\sigma_c$, the total time for a screw dislocation line to move to the adjacent Peierls energy dip increases by the time required for
a kink to escape from the traps of each H atom 
behind the dislocation.

We assume that the average time of a single escape event follows the Arrhenius law with the same prefactor $D_d$ of the kink nucleation as $\exp(E_t/k_BT)/D_d W_k$.
The total escape time is then given by $ C_b N_d \exp(E_t/k_B T)/D_d W_k$. 
When this time is comparable to the average nucleation time without hydrogen given by $\exp(E_{K}/kB T)/N_d D_d$, the effect of the enhanced nucleation rate is completely negated.
Thus a  condition 
\beq
\frac{ C_b N_d \exp (E_t/k_B T)}{D_d W_k} <
 \frac{\exp(E_{K}/k_BT)}{N_d D_d}
\label{equ:sigmau2}
\eeq
must be satisfied, which is simplified to $E_{K} - E_t < k_B T \log(C_b N_d^2/W_k)$. 
If we use the experimental values of $E_{K}$, the difference $E_{K} - E_t$ decreases as the shear stress is increased, and there is a second upper critical stress $\sigma_{U2}$ above which Eq. (\ref{equ:sigmau2}) is not satisfied.

When the hydrogen  concentration at the E1/E2 basin is $C_b$, the enthalpy increase due to the hydrogen de-trapping per unit length of kink motion is $C_b E_t^0 /b$, where $E_t^0=179$ meV.
If this is larger than the enthalpy gain $-\sigma b h$, the kink motion becomes impossible.
Thus there is a lower critical stress $\sigma_L = C_b E_t^0 /hb^2$ below which the screw dislocation cannot move.

In total, there are four conditions which must be satisfied for the enhanced  screw dislocation mobility of a screw dislocation, summarized as follows:
\begin{enumerate}
\item The temperature must be below $T_U$, above which hydrogen concentration at the promotion site is insufficient for the enhanced kink nucleation. $T_U$ increases as $C_0$ increases.
The boundary is given by $C_b^E(T_U)=C_b(C_0, T_U)$.
\item The applied shear stress must be greater than $\sigma_L$, below which a screw dislocation cannot move owing to the kink trapping by hydrogen.
$\sigma_L$ increases as $C_0$ increases.
The boundary is given by $\sigma_L=C_b(C_0,T)E_t^0/hb^2$.
\item The applied shear stress must be lower than $\sigma_{U1}$, above which dislocation motion is too fast and hydrogen concentration cannot catch up with it.
$\sigma_{U1}$ is independent of $C_0$.
The boundary is given by $E_{KH}(\sigma_{U1})=E_H^m$.
\item The applied shear stress must be lower than $\sigma_{U2}$, above which the time required for a kink to escape from H atoms behind the dislocation exceeds the average dislocation migration time without the H atom.
$\sigma_{U2}$ decreases as $C_0$ increases or the length of the screw dislocation increases.
The boundary is given by $E_{K}(\sigma_{U2})-E_t(\sigma_{U2})=k_B T \log(C_b(C_0,T)N_d^2/W_k)$.
\end{enumerate}
Fig. \ref{fig170sigu1} shows these conditions in the stress-temperature diagram for the two cases (a) $C_0=0.1$ appm and (b) $C_0=10$ appm, together with the experimentally observed yield stress compiled in Ref. \cite{caillard10}.
In each figure, plots of $\sigma_{U2}$ for two typical dislocation length, $2.0$ and $0.2 \mu$m are shown.
For the sake of comparison with the experiments, values of $E_K$ are taken from the experiment to evaluate the four conditions.
If we use the DFT values instead, the boundary of $\sigma_{U1}$ and $\sigma_{U2}$ shifts to the right side.
As can be seen in Fig. \ref{fig170sigu1}, $\sigma_L$ abruptly increases at the temperature where $C_b$ becomes of the order of $0.1$, and exceeds the yield stress.
Thus it is more appropriate to regard the boundary marked by  $\sigma_L$ as a lower critical temperature.

This result qualitatively agrees with the experimental observation of H-induced flow stress reduction in the macroscopic samples by Matsui and Kitamura \cite{matsui79}.
In their experiments, the bulk hydrogen concentration was estimated to be (at least) $15-30$ appm.
Reduction of the flow stress by the hydrogen charging was observed, and this reduction was more significant at lower temperatures (as is the general case of the solution softening), whereas below $190$ K, hardening caused by hydrogen charging was observed.
The maximum flow stress in their experiment was $200$ MPa, and thus the shear stress in any slip plane does not exceed $100$ MPa because the  Schmid factor is always less than $1/2$.
The transition at $190$ K can be attributed to the crossing of $\sigma_L$, although for $C_0=30$ appm the crossing occurs at a much higher temperature in our model.
The quantitative disagreement of the transition temperature can be attributed to either the numerical uncertainty of the binding energy, or to our approximation of the hydrogen concentration at the binding site to the equilibrium concentration.
The actual hydrogen concentration behind the moving dislocation must be much smaller, and in that case the transition temperature decreases.


\section{Summary and Conclusion}
Using DFT calculations with careful estimations of the finite-size effects, we have calculated the binding energies of an H atom at various binding sites around a screw dislocation  of bcc Fe for two kinds of core configurations.
The strongest binding energy to the stable and unstable core configurations were estimated as $256\pm32$ and $390\pm32$ meV, respectively.
Experimentally observed hydrogen binding energy to the screw dislocations is currently not available, because in thermal desorption spectroscopy the desorption peak for screw dislocations is insignificant compared to that of edge dislocations and grain boundaries.

The interaction between an H atom and a screw dislocation was incorporated into a line tension model of a curved dislocation line.
Using this model, the reduction of kink nucleation enthalpy by hydrogen was estimated to be $110$ meV.
The softening effect of H atoms by promoting kink nucleation and the hardening effect by trapping the kink movement were both evaluated, and four conditions for the overall softening were derived.
These conditions consist of two upper critical stresses and upper/lower critical temperatures.
The upper critical stresses were found to be far above the yield stress, and only relevant for very high strain rate cases.
The temperature range for the softening bounded by the upper/lower critical temperatures was roughly estimated to be $200-300$ K for the bulk hydrogen concentration of $0.1$ appm case and $300-400$ K for $10$ appm case.
For a precise estimate of the temperature range, a kinetic Monte Carlo simulation which incorporates both the dislocation motion and the hydrogen diffusion is required.
A clear transition between softening and hardening behavior at the lower critical temperature is predicted, which is in qualitative agreement with experimental observations \cite{matsui79}.

While the dislocation migration behavior in ferritic steels is very different compared to the pure iron case owing to the presence of carbon and other solute atoms,  the low-temperature hardening caused by the dense hydrogen segregation to the screw dislocation is expected to be independent of other solute atoms.
It is expected that properties concerning  the softening to hardening transition at low temperatures are observed for various hydrogen concentrations, strain rates and material purities.

\begin{table*}[htb]
\renewcommand{\arraystretch}{1.2} 

\caption{
\label{tab:energy}
Number of $k$-points $N_k$, solution energy $E_s$,
binding energy $E_b$, ZPE correction $E_{z}$,
ZPE correction to the binding energy $\Delta E_{z}$,
total binding energy, and estimated
numerical and finite-size error for each binding site shown in Fig. \ref{fig00bind}.
The ``00'' case is the reference values of perfect crystal t-site.
All energies are in meV.
}

\begin{tabular}{ccccccccc}
Site & $N_k$ & $E_s$ & $ E_b$ & $E_{Z}$ & $\Delta E_{Z}$ & $E_s^Z$ & $E_b^Z$ & Err\\
 \hline
 00 & $1\times 1\times 8$ & $241$ & $  0$ & $238$ &  $0$ & $346$ & $0$   & $\pm 57$\\ 
    & $3\times 3\times 8$ & $215$ & $  0$ & $238$ &  $0$ & $320$ & $0$   & $\pm 32$\\ 
 \hline
 E0 & $1\times 1\times 8$ & $156$ & $ 85$ & $224$ & $ 14$ & $247$ & $99$  & $\pm 57$ \\ 
 E1 & $3\times 3\times 8$ & $22 $ & $193$ & $182$ & $ 56$ & $ 71$ & $249$ & $\pm 32$ \\ 
 E2 & $3\times 3\times 8$ & $30 $ & $185$ & $167$ & $ 71$ & $ 64$ & $256$ & $\pm 32$ \\ 
 E3 & $1\times 1\times 8$ & $103$ & $138$ & $175$ & $ 63$ & $145$ & $201$ & $\pm 57$ \\ 
 E4 & $1\times 1\times 8$ & $119$ & $122$ & $176$ & $ 62$ & $162$ & $184$ & $\pm 57$ \\ 
 E5 & $1\times 1\times 8$ & $211$ & $ 30$ & $245$ & $ -7$ & $323$ & $ 23$ & $\pm 57$ \\ 
 E6 & $1\times 1\times 8$ & $231$ & $ 10$ & $248$ & $-10$ & $346$ & $  0$ & $\pm 57$ \\ 
 E7 & $1\times 1\times 8$ & $247$ & $ -6$ & $234$ & $  4$ & $348$ & $ -2$ & $\pm 57$ \\ 
 E8 & $1\times 1\times 8$ & $195$ & $ 46$ & $207$ & $ 31$ & $269$ & $ 77$ & $\pm 57$ \\ 
 E9 & $1\times 1\times 8$ & $236$ & $  5$ & $225$ & $ 13$ & $328$ & $ 18$ & $\pm 57$ \\ 
 \hline
 H0 & $3\times 3\times 8$ & $-71$ & $286$ & $134$ & $104$ & $-70$ & $390$ & $\pm 32$ \\ 
 H1 & $3\times 3\times 8$ & $-56$ & $271$ & $187$ & $ 51$ & $ -2$ & $322$ & $\pm 32$ \\ 
 H2 & $1\times 1\times 8$ & $116$ & $125$ & $174$ & $64 $ & $157$ & $189$ & $\pm 57$ \\ 
 H3 & $1\times 1\times 8$ & $183$ & $ 58$ & $249$ & $-11$ & $299$ & $ 47$ & $\pm 57$ \\ 
 H4 & $1\times 1\times 8$ & $218$ & $ 23$ & $229$ & $9$   & $314$ & $ 32$ & $\pm 57$ \\ 
 H5 & $1\times 1\times 8$ & $236$ & $  5$ & $222$ & $16 $ & $325$ & $ 21$ & $\pm 57$ \\ 
 H6 & $1\times 1\times 8$ & $247$ & $ -6$ & $227$ & $11 $ & $341$ & $  5$ & $\pm 57$ \\ 
\hline
\end{tabular}

\end{table*}

\myfig{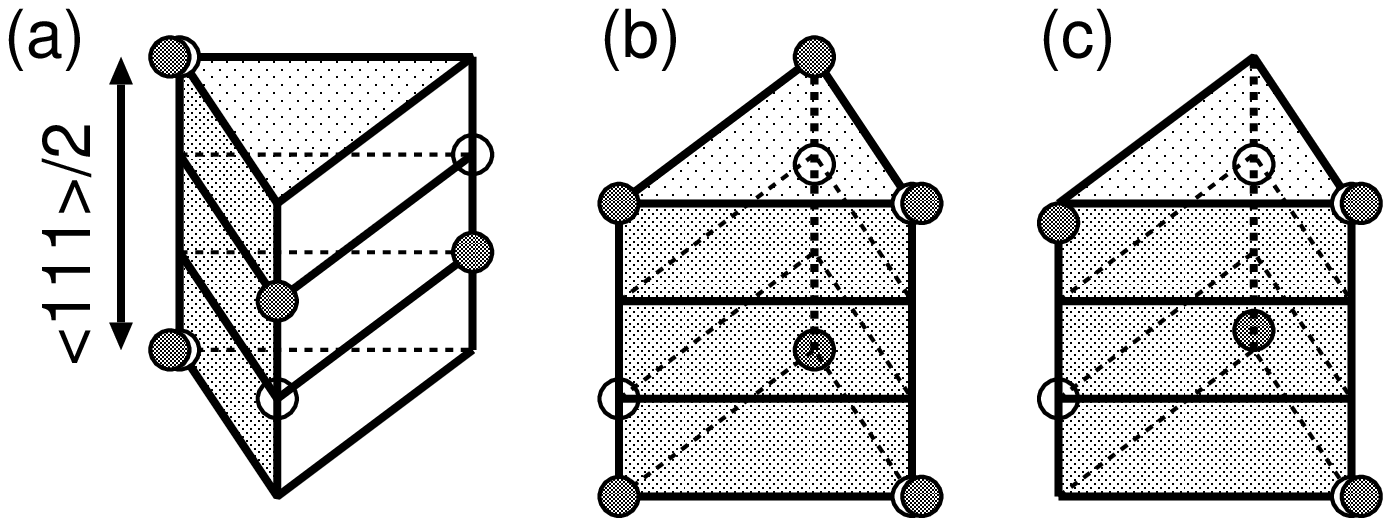}
{Atomic structures of a screw dislocation in the (a) easy core, (b) hard core and (c) migration saddle point configurations, shown by gray spheres. White spheres are the atom positions in a perfect bcc crystal.}

\myfig{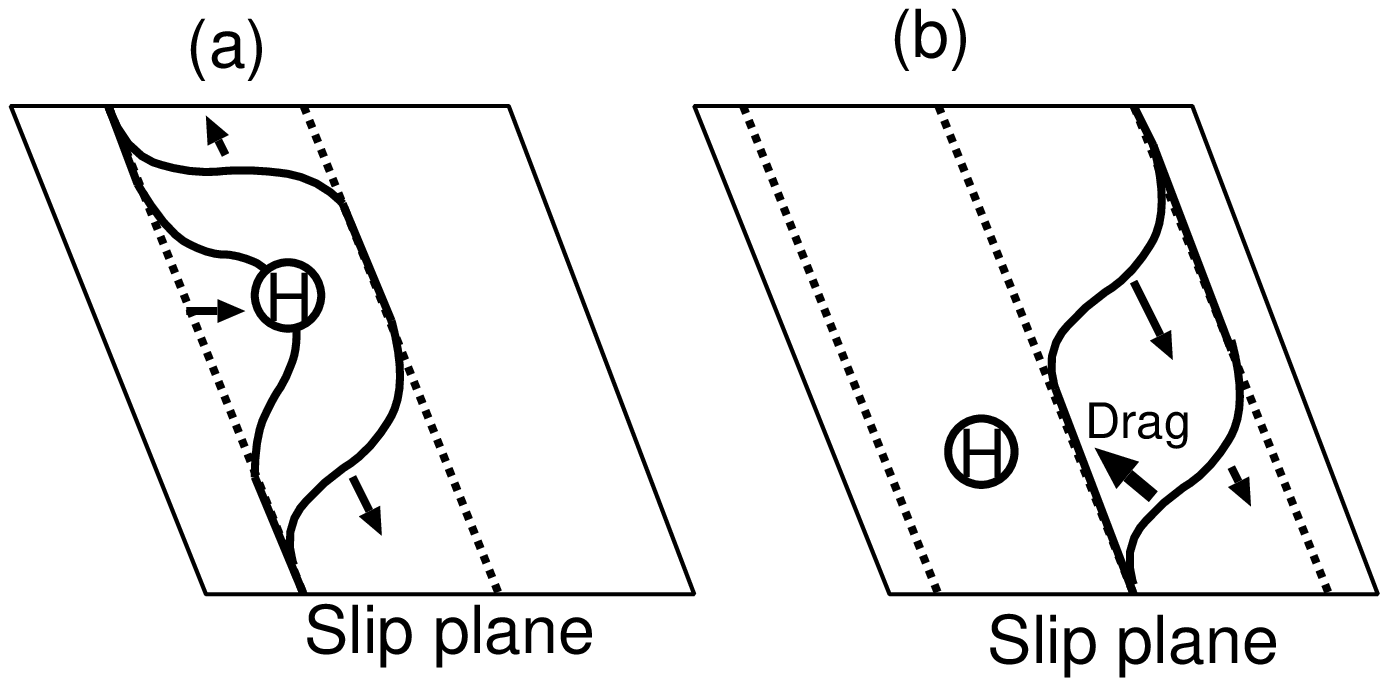}
{ Schematic diagram of the effects of an H atom on the migration process of a screw dislocation by (a) lowering the Peierls barrier and (b) decreasing the kink velocity. See the main text for details.  }

\myfig{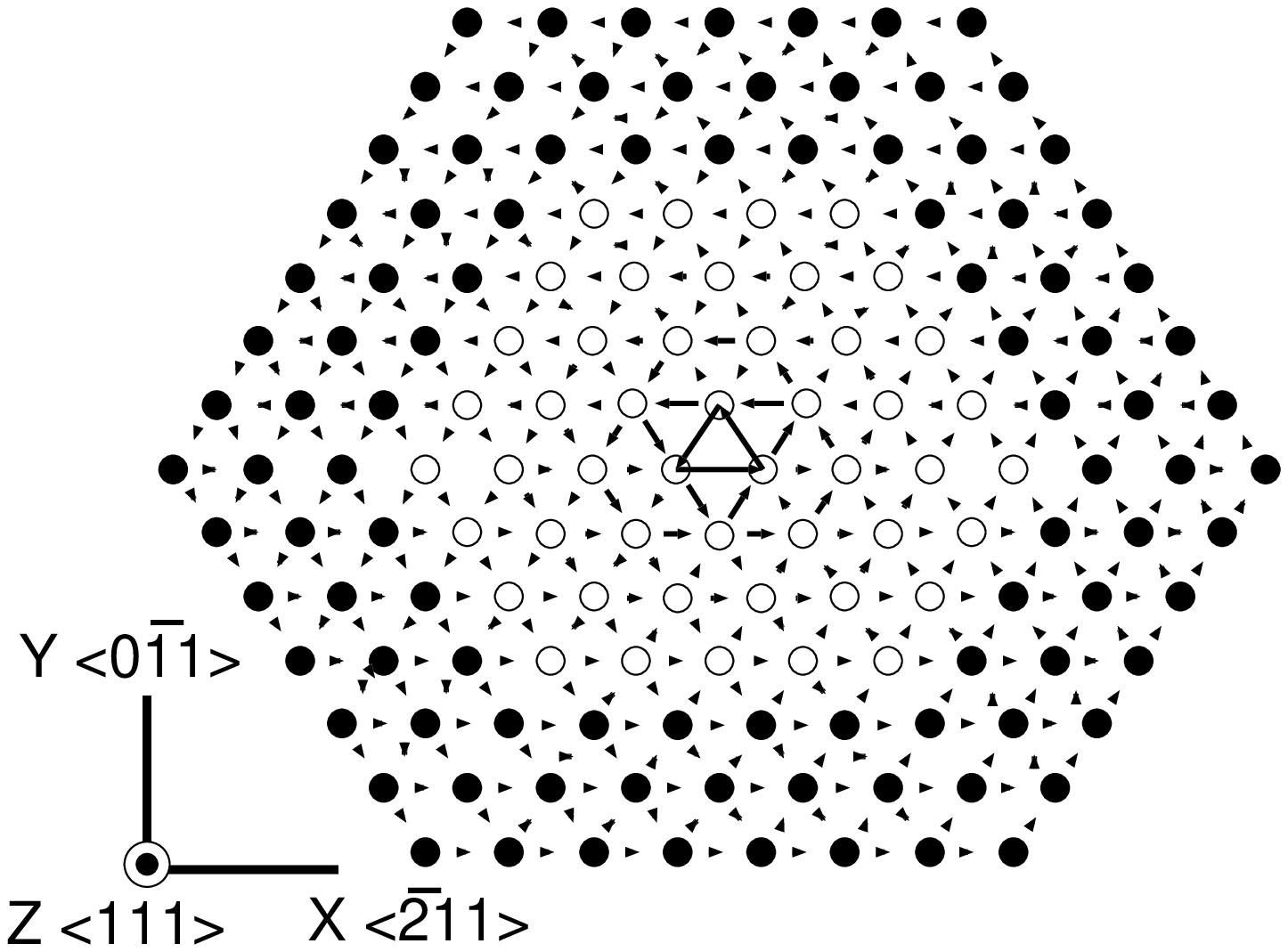} {Two regions 1 and 2 used for the flexible boundary method in DFT,
shown by white and black circles, respectively.
The arrows show the differentiated displacement of the easy core configuration.
The Cartesian coordinates used throughout the present paper are also shown.}

\myfig{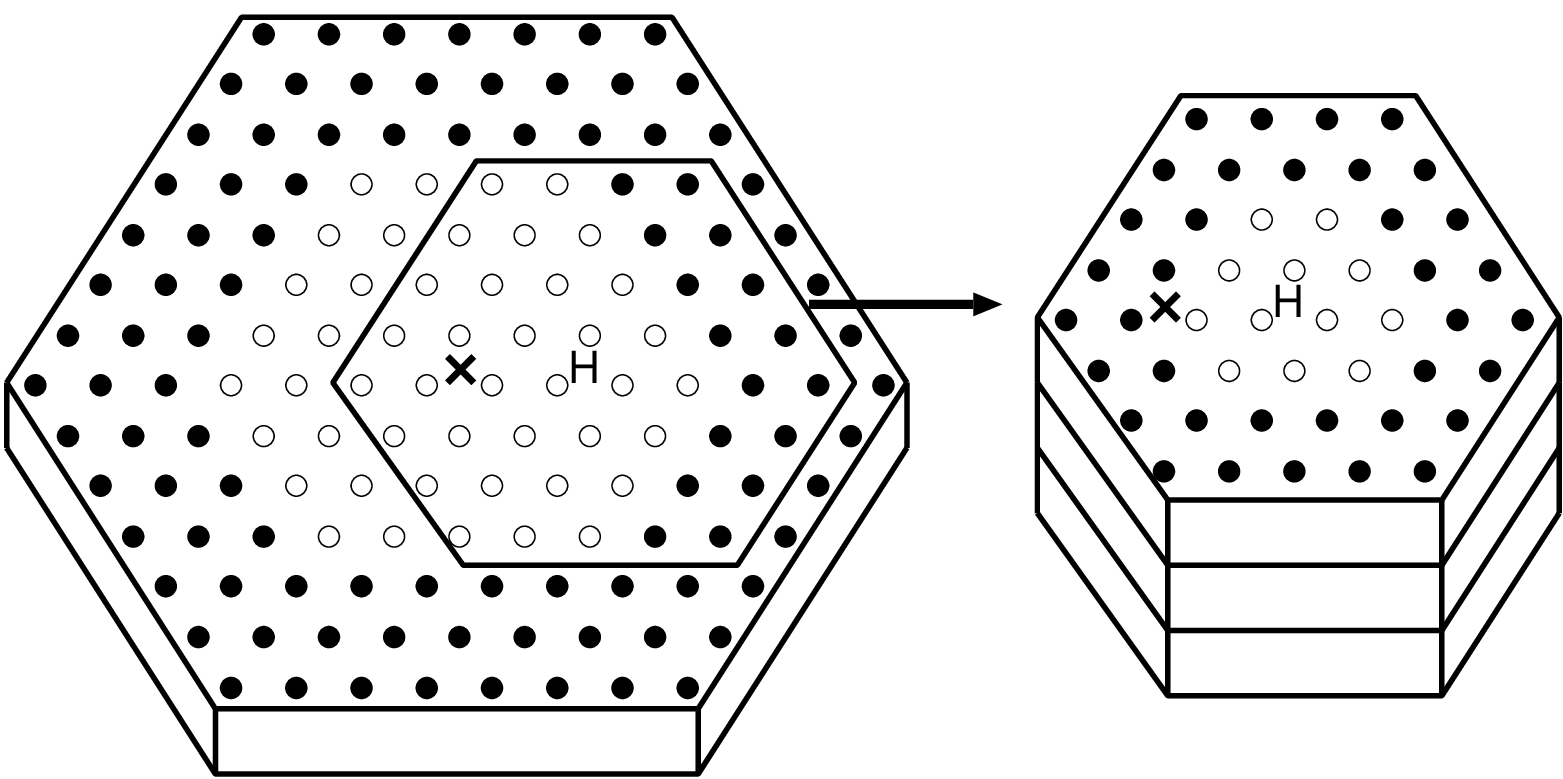} {
Calculation cell of the reference dislocation configuration (left)
and dislocation$+$H configuration (right).
Fe atoms 
close to the hydrogen trap site are clipped out from the reference configuration
and replicated in the $Z$ direction three times, H atom is placed at the trap site,
and structural relaxation is then performed.
The cross symbol, white and black circles denote the core position,
and movable and fixed Fe atoms, respectively. }

\myfigv{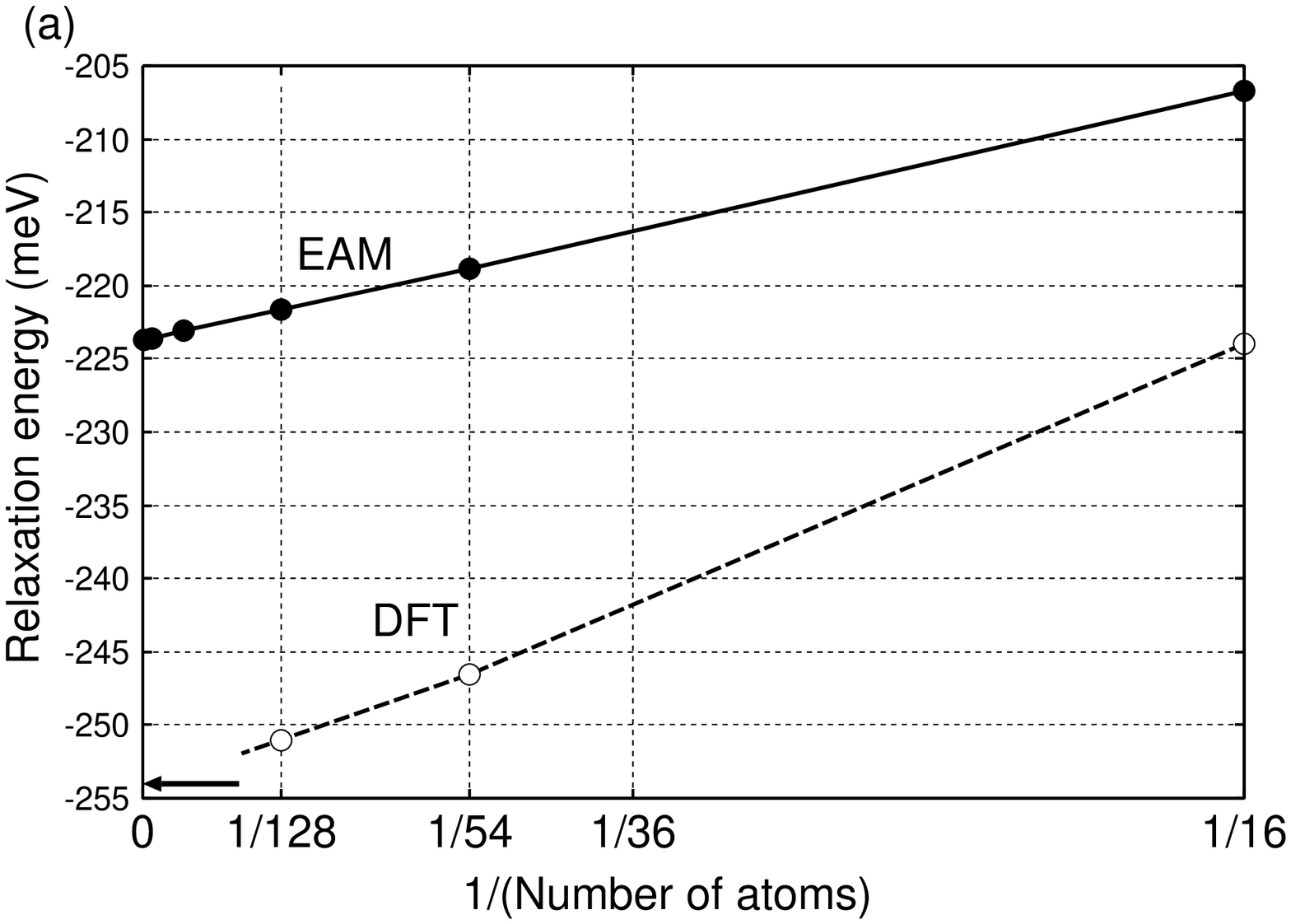}{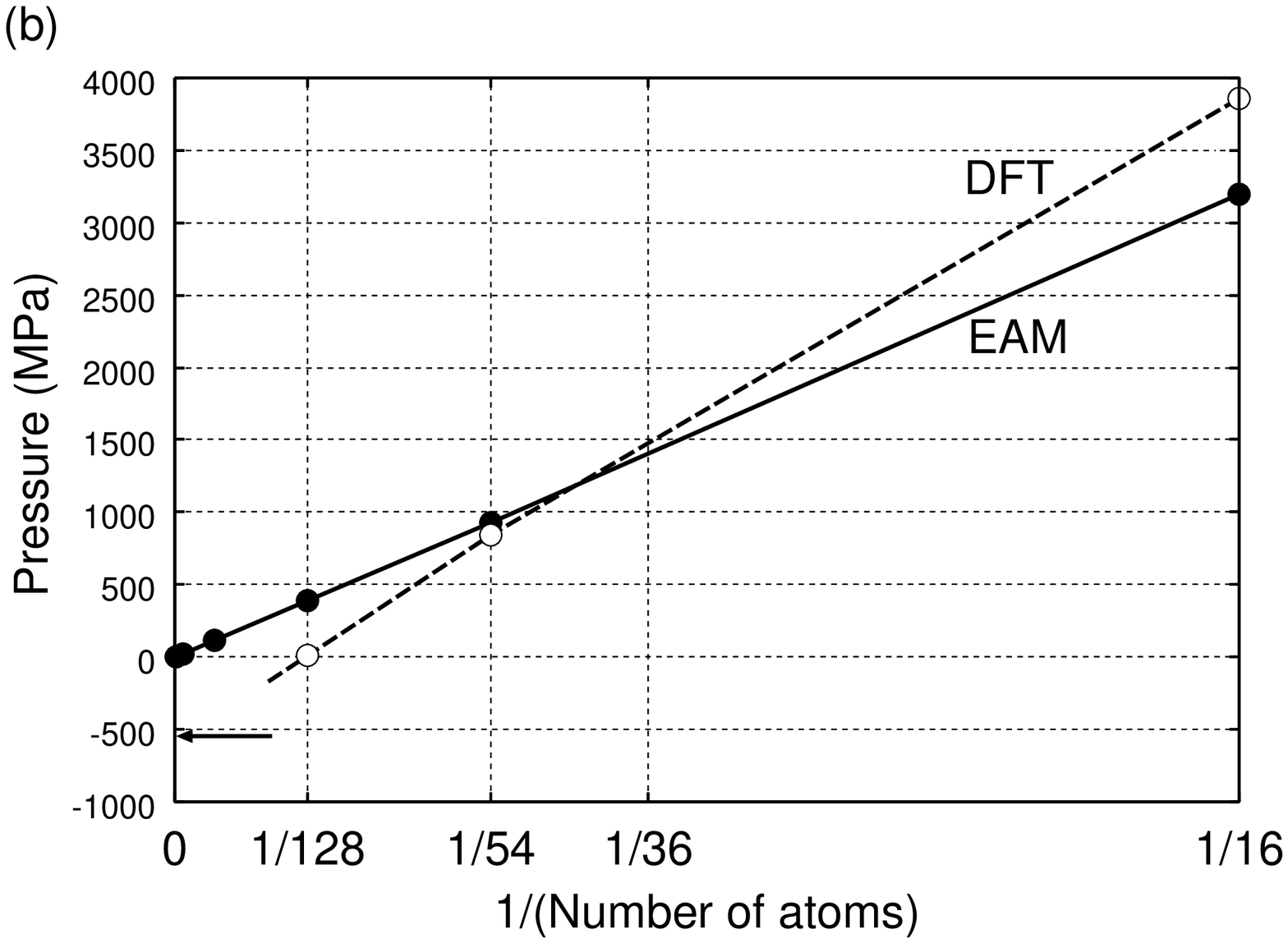}
{Size dependence of (a) relaxation energy and (b) pressure in the system of a bcc Fe perfect crystal with an H atom at the t-site under constant-volume boundary conditions.  The result of the EAM potential \cite{kimizuka11} calculations is also shown.}

\myfig{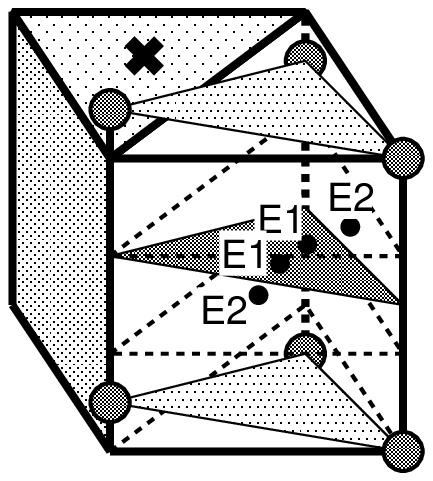}
{
Configuration of Fe atoms adjacent to the easy core. The cross marks the core position.
The binding sites E1 and E2 are on the midplane of the slanted triangular prism shown by
the dark shaded triangle.
}

\myfig{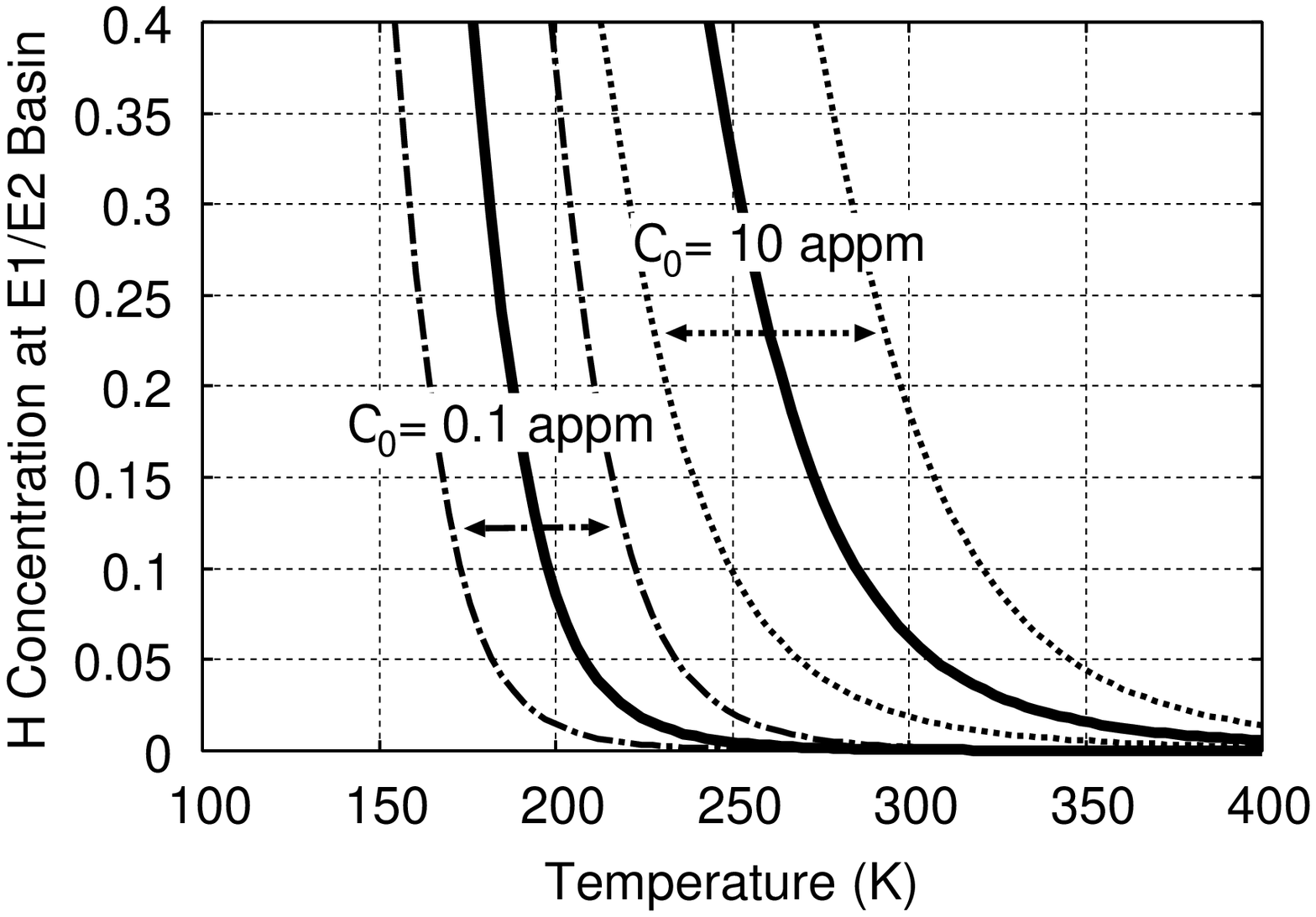}
{
H concentration at the binding site close to the easy core screw dislocation,
calculated by McLean's equation with binding energy $256\pm32$ meV for
two cases of bulk hydrogen concentrations.
The dashed and dotted lines show the range of numerical uncertainty.
}


\myfig{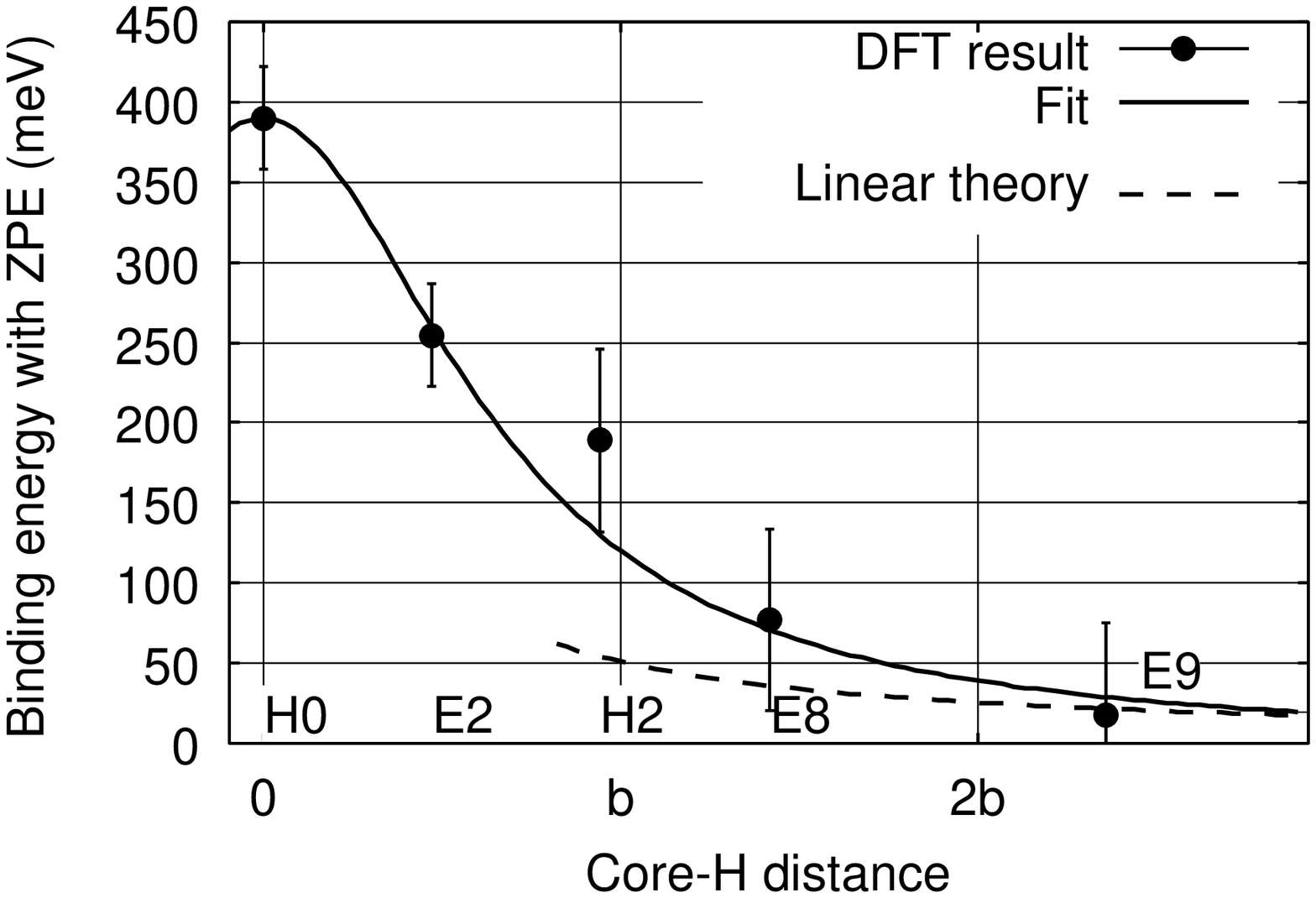}
{
Hydrogen binding energy plotted against the distance between an H atom
and dislocation core. A fitting curve to the DFT data is also shown by the
solid line. The dashed line is the binding energy derived by
linear elasticity theory \cite{wang13}.
}

\myfig{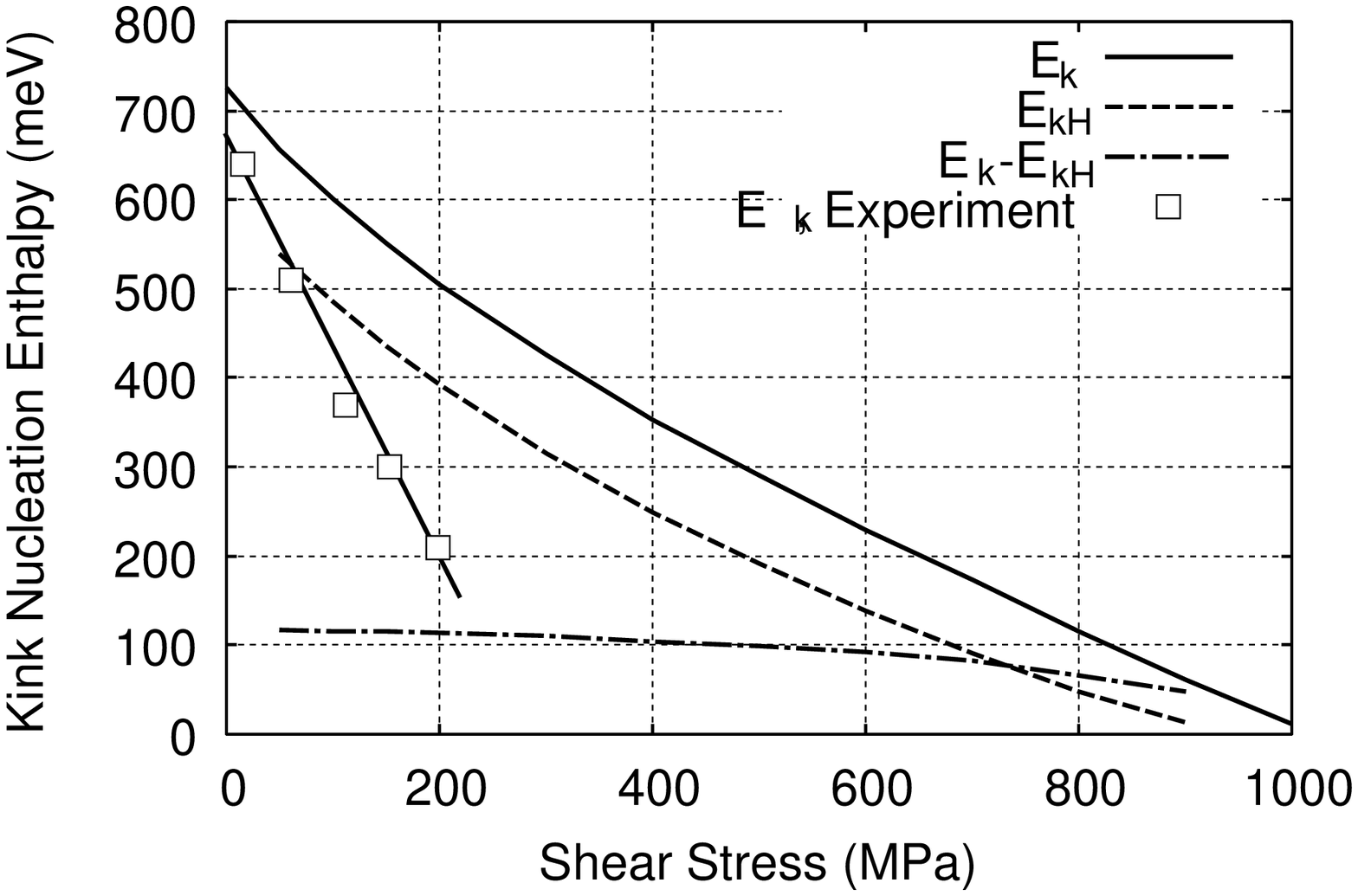}
{
Kink nucleation enthalpy with H ($E_{KH}$)
and without H ($E_{K}$) for the shear stress applied in
the $[111](1\bar{1}0)$ direction.
Estimated values of $E_{K}$ from the experiment \cite{spitzig70} are also
shown with a linear fitting line.
}

\myfig{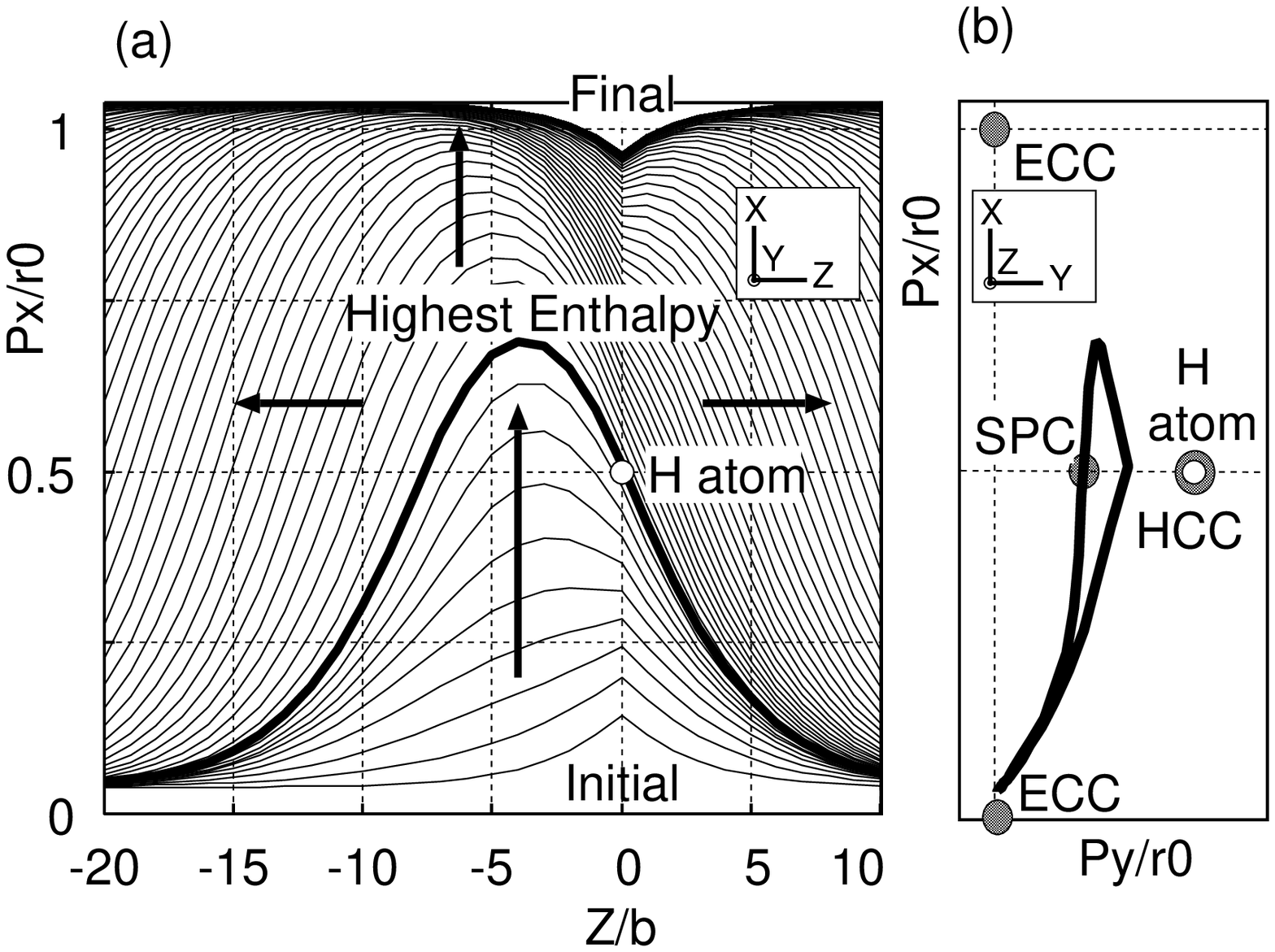}
{
(a)
Kink nucleation and migration process for $400$ MPa case in the presence of an H atom
shown by the white circle.  The bold line represents the highest enthalpy configuration.
(b)
The highest enthalpy configuration seen from the $Z$ direction.
The positions of the easy core, the hard core and the saddle point of the Peierls barrier are
also shown.
}


\myfig{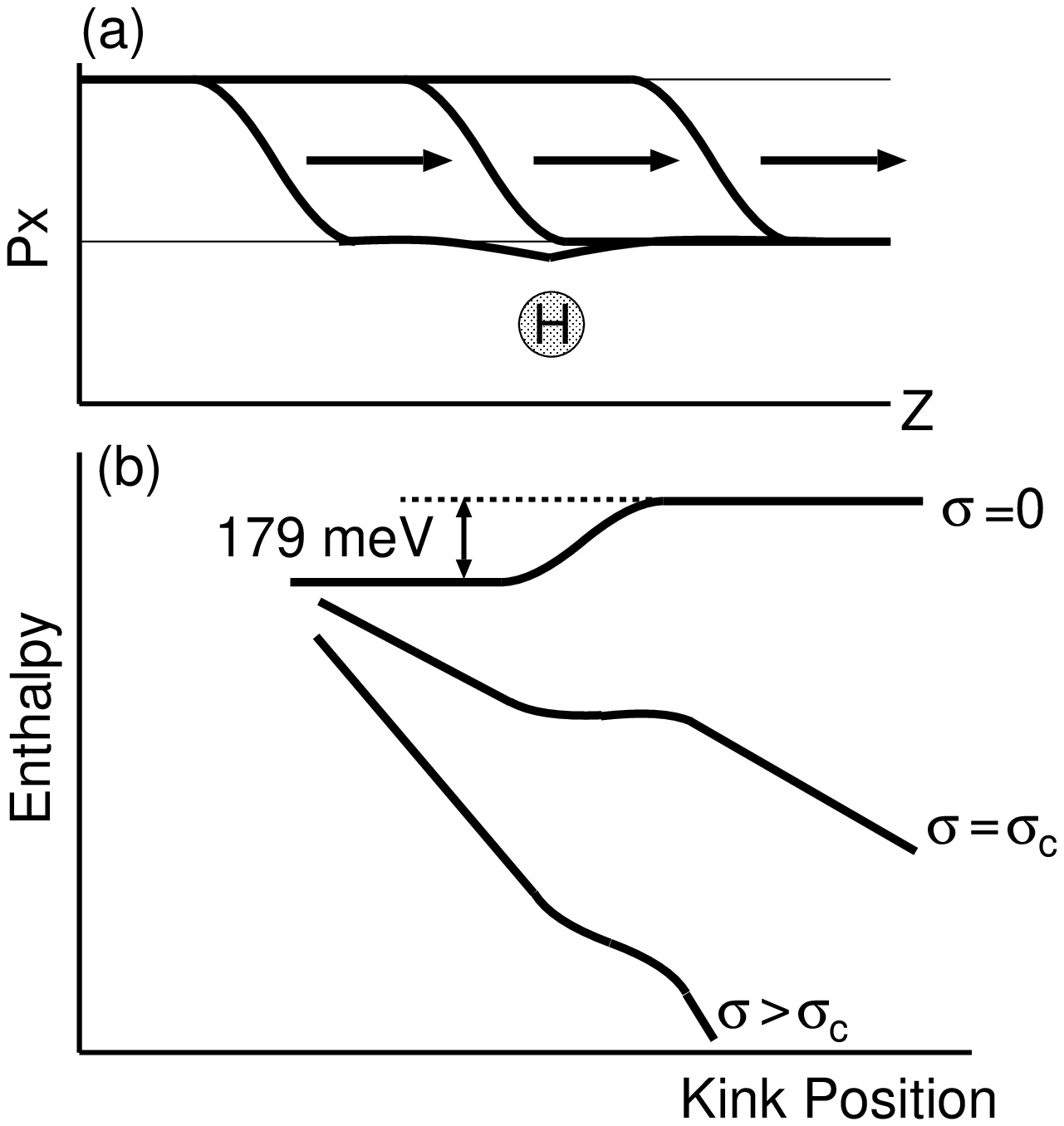}
{ Schematic of kink trapping effect by the H atom.
(a) As a kink moves from left to right and passes the H atom
behind the dislocation line, the binding between the dislocation
and the H atom is weakened.
(b) The enthalpy increases by the amount of weakened binding energy.
As the shear stress is applied,
a slope proportional to the stress is added to the enthalpy curve
and at some critical stress $\sigma_c$ the barrier vanishes.
}

\myfig{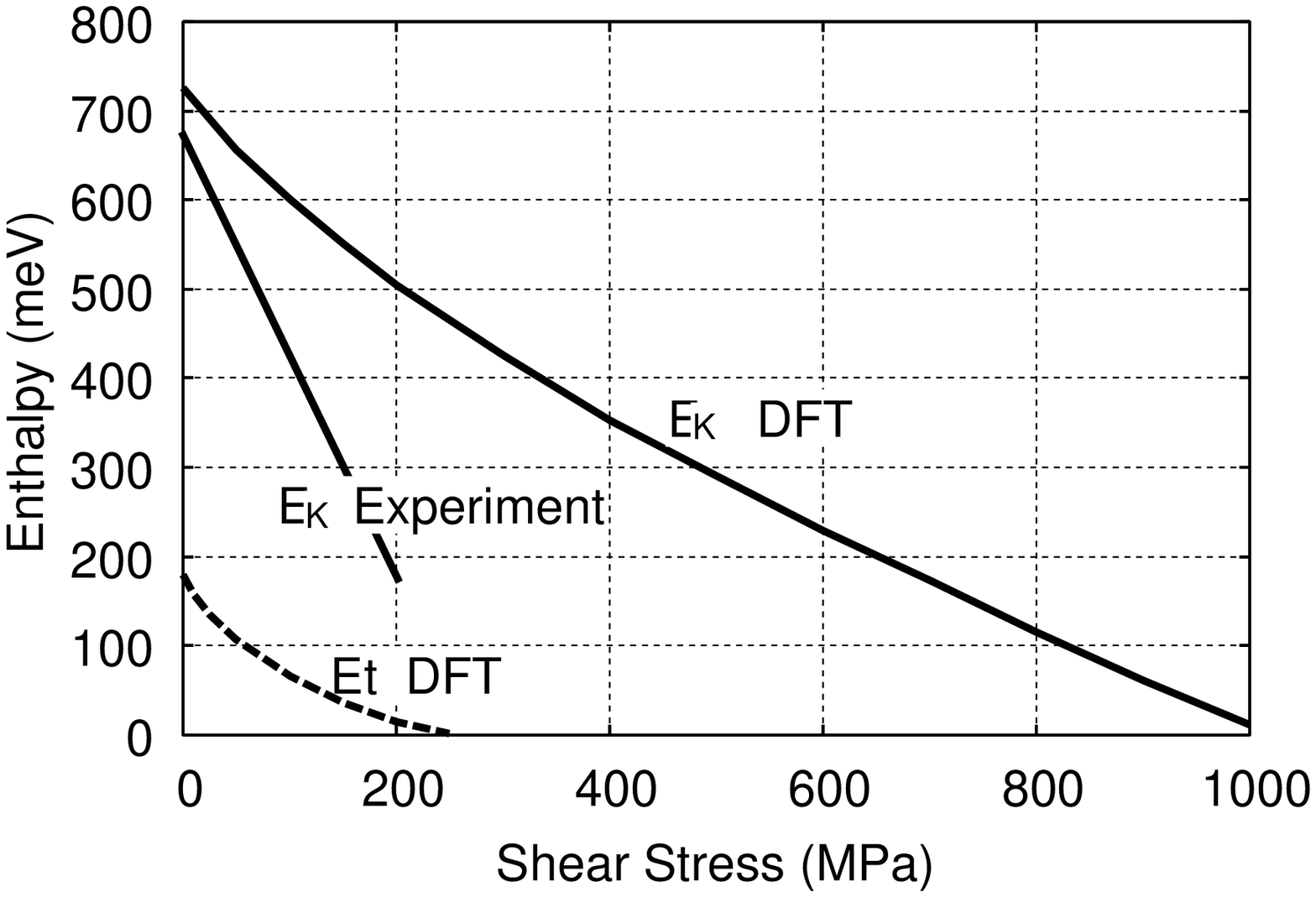}
{
Kink trap enthalpy $E_t$ plotted against the shear stress.
Plots of the kink nucleation enthalpy $E_K$ are also shown.
}

\myfigv{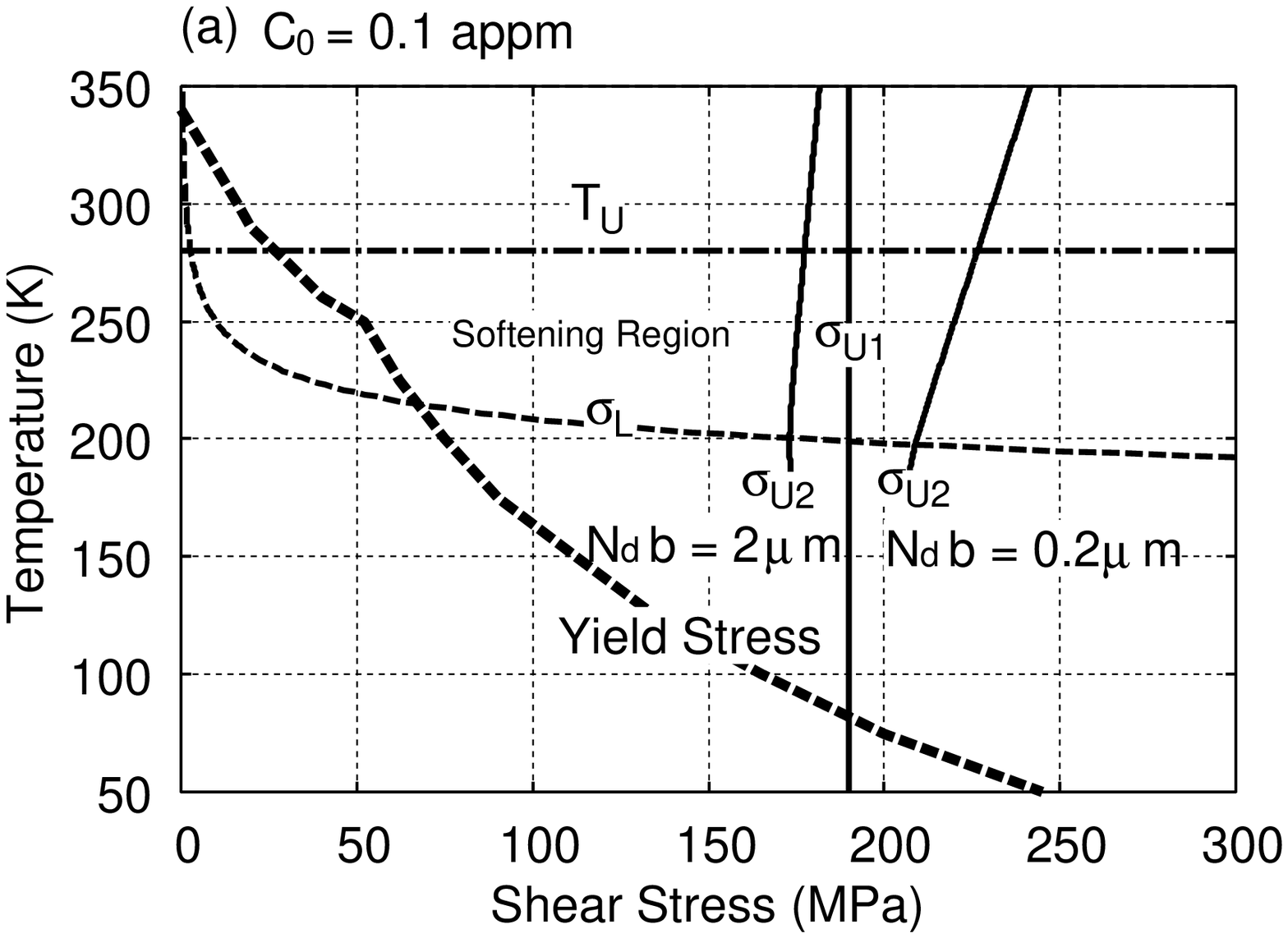}{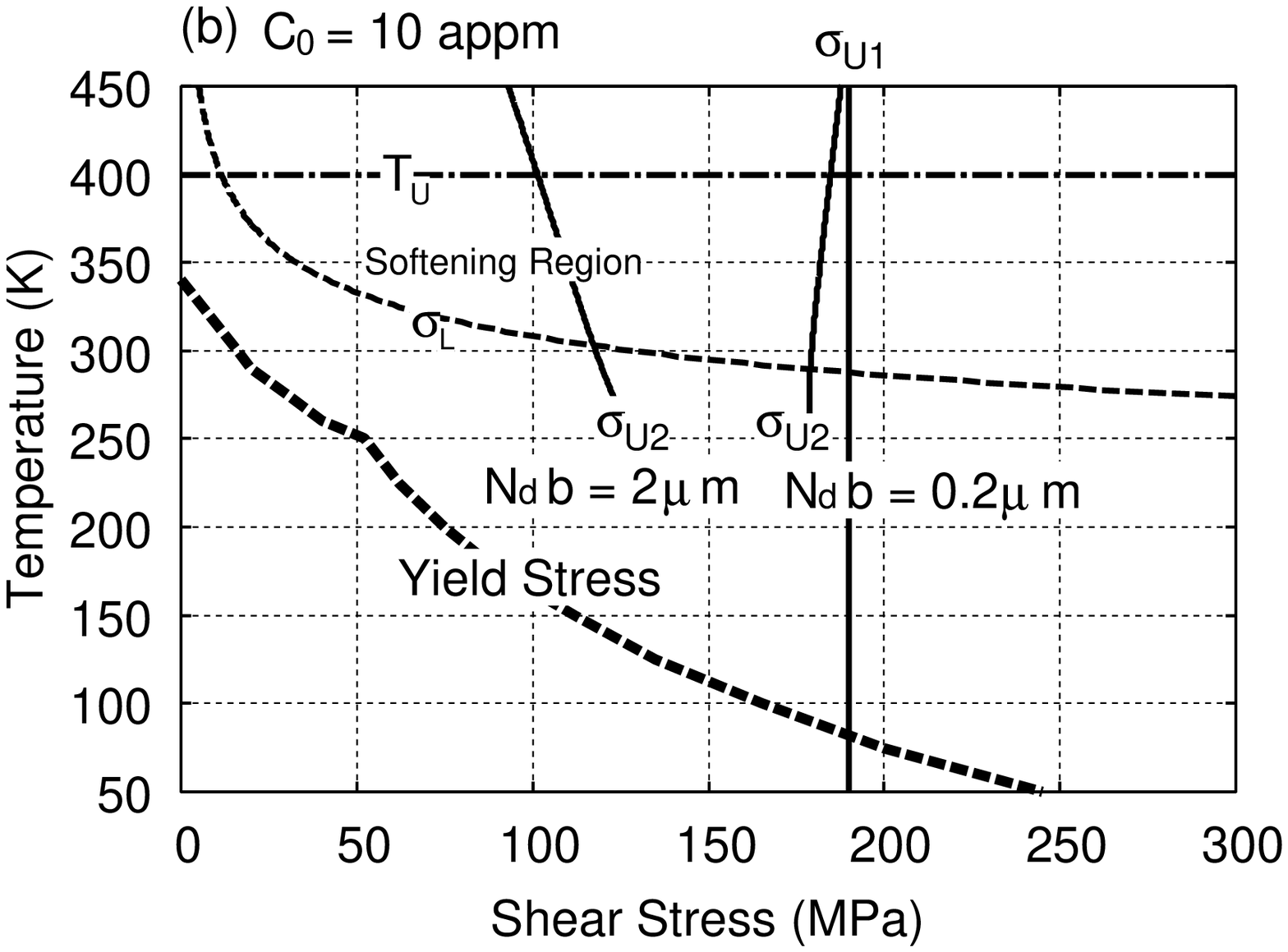}
{
A region in the temperature-stress diagram where 
an increase in the screw dislocation velocity by hydrogen solution
is possible for the bulk hydrogen concentration of (a) $0.1$ at.ppm
and (b) $10$ at.ppm.
The region is bounded by the upper critical temperature $T_U$,
the lower critical stress $\sigma_L$, the yield stress, and two kinds of 
upper critical stresses $\sigma_{U1}$ and $\sigma_{U2}$.
$\sigma_{U2}$ depends on the dislocation length, and 
two cases of typical dislocation length are shown in each figure.
}

\begin{thebibliography}{99}
\bibitem{aniso}
The anisotropic
and isotropic linear elastic solutions
of the displacement field of the screw dislocation are identical.
See Ref. \cite{itakura12}.
\end{thebibliography}
\end{document}